\documentclass[a4paper]{mem}
\usepackage{natbib}
\usepackage{graphicx}
\usepackage[a4paper]{hyperref}
\idline{73}{1}
\begin{document}
   \title{Rotation and Disk Accretion in Very Low Mass Stars and Brown Dwarfs
}

   \author{Jochen Eisl\"offel \inst{1} 
          \and
          Alexander Scholz \inst{1}
}

   \offprints{Jochen Eisl\"offel}

   \institute{Th\"uringer Landessternwarte, Sternwarte 5, D-07778 Tautenburg, Germany}

   \abstract{
The regulation of angular momentum is one of the key processes for our 
understanding of stellar evolution. In contrast to solar-mass stars, very 
low mass (VLM) objects and brown dwarfs are believed to be fully
convective. This may lead to major differences of rotation and activity, since
fully convective objects may not host a solar-type dynamo.
Here, we report on our observational efforts to understand the rotational 
evolution of VLM objects. 
   \keywords{Stars: activity, evolution, formation, low-mass, brown dwarfs, 
late-type, rotation}
   }
   \authorrunning{J. Eisl\"offel and A. Scholz}
   \titlerunning{Rotation and Disk Accretion in VLM Objects}
   \maketitle
%

\section{Introduction}

Rotation is one of the key parameters for stellar evolution. It is the 
parameter that -- at least in some well-behaved objects -- can be measured 
to the highest accuracy. A precision for the rotation periods of 1 : 10000 
is possible. 

In solar mass stars the investigation of their rotation has allowed us new
insights into their evolution (\citealt{b95}, \citealt{bfa97}, 
\citealt{st03}). It has become clear that angular momentum
regulation is a direct consequence of basic stellar physics: most of the
angular momentum of a fragmenting and collapsing molecular cloud is lost in
the course of the formation of protostars. The specific angular momentum of 
protostars is, however, still one or two orders of magnitude higher than 
that of young main sequence stars. On the other hand, in their T\,Tauri 
phase solar-mass stars rotate slowly although they are accreting. The magnetic
coupling between the star and its disk, and consequent angular momentum
removal in a highly collimated bipolar jet are thought to be responsible for
this rotational braking (\citealt{c90}, \citealt{k91}, \citealt{snowrl94}). 
After the dispersal of the disk, and thus loss of the
braking mechanism, the rotation is observed to accelerate as the stars
contract towards the zero-age main sequence. On the main sequence 
the rotation rates of solar-mass stars decrease again because of angular
momentum loss through stellar winds.

Rotation can be investigated either by measuring stellar photospheric lines
spectroscopically, or by the determination of rotation periods from
photometric time series observations. While the former suffers from projection
effects -- the unknown inclination angle of the rotation axis with respect to
the line of sight -- the latter can be determined with high precision and
independent of inclination angle.
For our monitoring programme we therefore decided to follow the photometric 
time series approach to obtain precise rotation periods.

\section{Observations and data analysis}

Since hardly any rotation periods for VLM objects and brown dwarfs with ages
older than 3\,Myr were known, it was necessary to create a database 
complementing the known rotation periods of solar-mass stars. 

\begin{figure}[t]
  \begin{center}
    \resizebox{\hsize}{!}{\includegraphics[width=7cm, angle=-90]{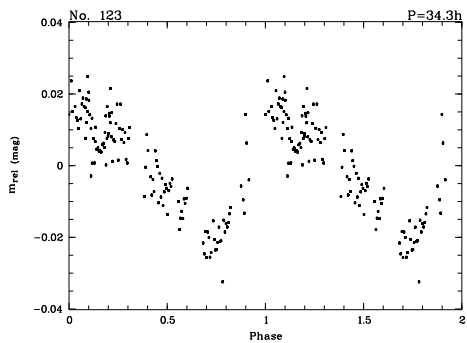}}
  \end{center}
\caption{Phase-folded light curve for a 50 M$_{jup}$ brown dwarf in the
  epsilon Ori field. The measured rotation period is 34.3\,h, and the light
  curve comprises of 129 measured data points. \label{fig1}}
\end{figure}
In the course of our ongoing monitoring programme we have so far obtained
rotation periods for 23 VLM objects in the cluster around \object{sigma Ori} 
(\citealt{se04a}), for 30 in the field around \object{epsilon Ori} 
(\citealt{se04c}), which are belonging to the Ori\,OBIb association, and for 
9 objects in the \object{Pleiades} open cluster (\citealt{se04b}). With ages of
about 3, 5, and 125\,Myr these three groups of VLM objects form an age
sequence that already allows us insights into a relevant part of their young
evolution. 

Our time series photometry was done with the Wide Field Imager 
(WFI) at the ESO/MPG 2.2-m telescope on La Silla in \object{epsilon Ori},
and with the CCD camera at the 1.23-m telescope at the German-Spanish
Astronomy Centre on Calar Alto (CA) in the \object{Pleiades}. The 
\object{sigma Ori} cluster was observed in two campaigns with the CCD 
cameras at the 2-m Schmidt telescope in Tautenburg (TLS) and at the 1.23-m 
telescope on Calar Alto. Our data analysis is described in full detail 
in \citet{se04a} and \citet{se04b}. Fig.\,1 shows an example for a final 
phase-folded light curve of a brown dwarf in the \object{epsilon Ori} field. 

\begin{figure*}[th]
  \begin{center}
    \resizebox{\hsize}{!}{\includegraphics[angle=-90]{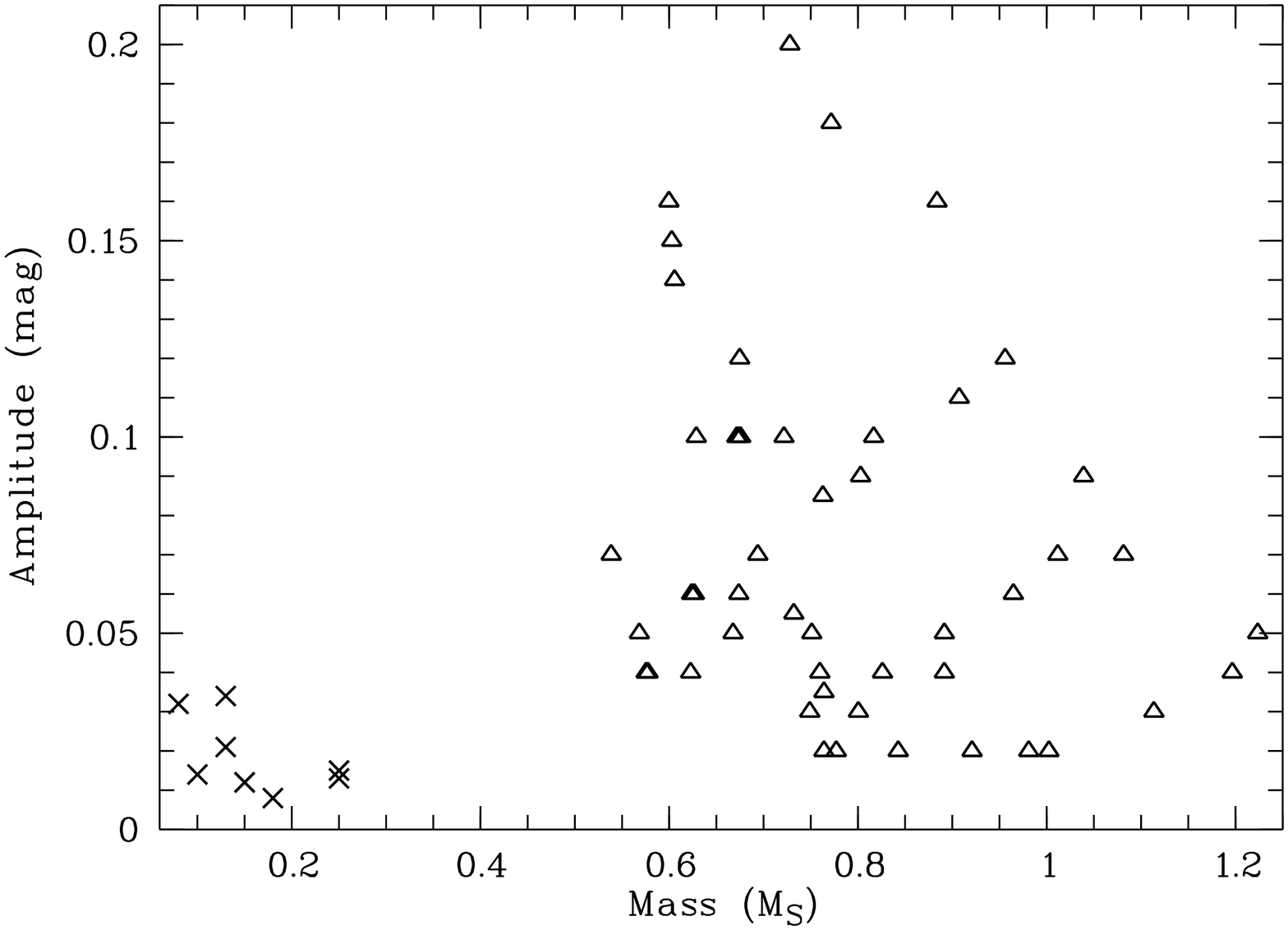}\includegraphics[angle=-90]{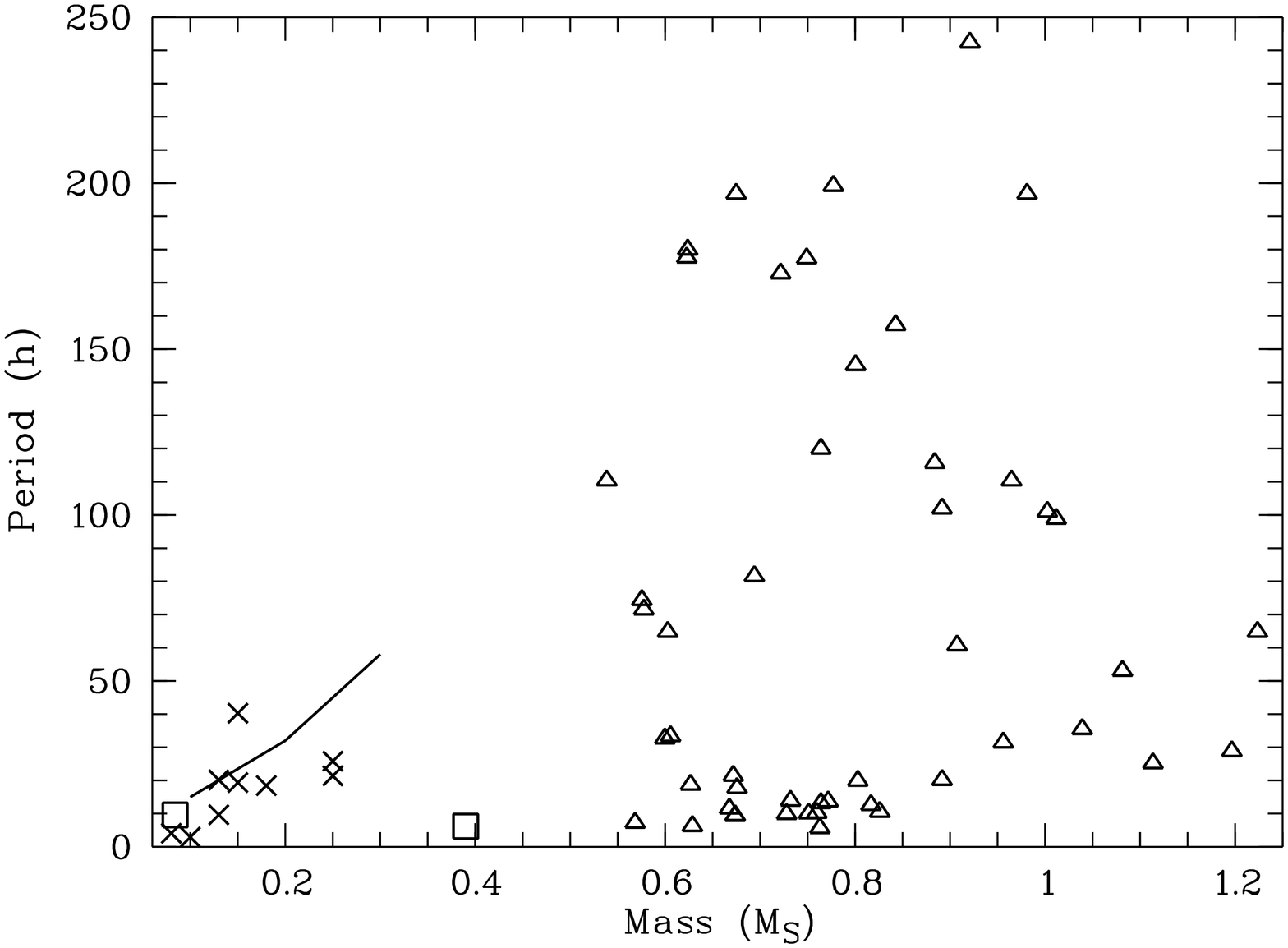}}
  \end{center}
\caption{Photometric amplitude versus mass for \object{Pleiades} stars from 
the Open Cluster Database (triangles, see Sect.\,3 for complete references) 
and our targets (crosses). The detection limit for the solar-mass stars is 
0.02\,mag, explaining the lack of stars with very low amplitudes in this 
sample.  \newline
Rotation periods versus mass in the \object{Pleiades}. Our rotation
 periods for VLM objects are shown as crosses. Triangles mark the periods for 
solar-mass stars from the Open Cluster Database. The two squares show 
periods from Terndrup et al. (1999) The solid line marks the upper 
limit to the observed $v \sin i$ values of Terndrup et al. (2000). 
\label{fig2}}
\end{figure*}

\section{Rotation and variability of VLM objects}

The general interpretation for the observed periodic variability in the light
curves of our VLM targets are surface features, which are asymmetrically
distributed on the surface and are co-rotating with the objects. Such surface
features could arise either from dust condensations in the form of ``clouds'',
or from magnetic activity in the form of cool ``spots''. Since all our objects,
because of their youth, have surface temperatures T$_{\rm eff}$ $>$ 2700\,K 
(\citealt{baraffe98}) corresponding to spectral types earlier than M8, and 
thus higher than the dust condensation limits, we are most likely observing 
the effects of cool, magnetically induced spots.

It is interesting to compare the photometric amplitudes of the periodic
variations in the light curves with those of more massive cluster
members. Such a comparison can be done for the \object{Pleiades}, for 
which the required photometric information for solar-mass stars is available 
from the Open Cluster Database (provided by C.F. Prosser (deceased) and 
J.R. Stauffer). 
Fig.\,2 shows that larger amplitude variations are only observed in the higher 
mass objects. It is statistically significant that the amplitude distributions 
for higher and lower mass objects are different. That only amplitudes smaller 
than 0.04\,mag are observed in the VLM objects may be attributed to the fact 
that a) the relative spot covered areas of their surfaces are smaller, b) 
their spot distributions are more symmetric or c) the spots have a lower 
relative temperature contrast with the average photosphere.

\begin{figure*}[ht]
  \begin{center}
    \resizebox{\hsize}{!}{\includegraphics[angle=-90]{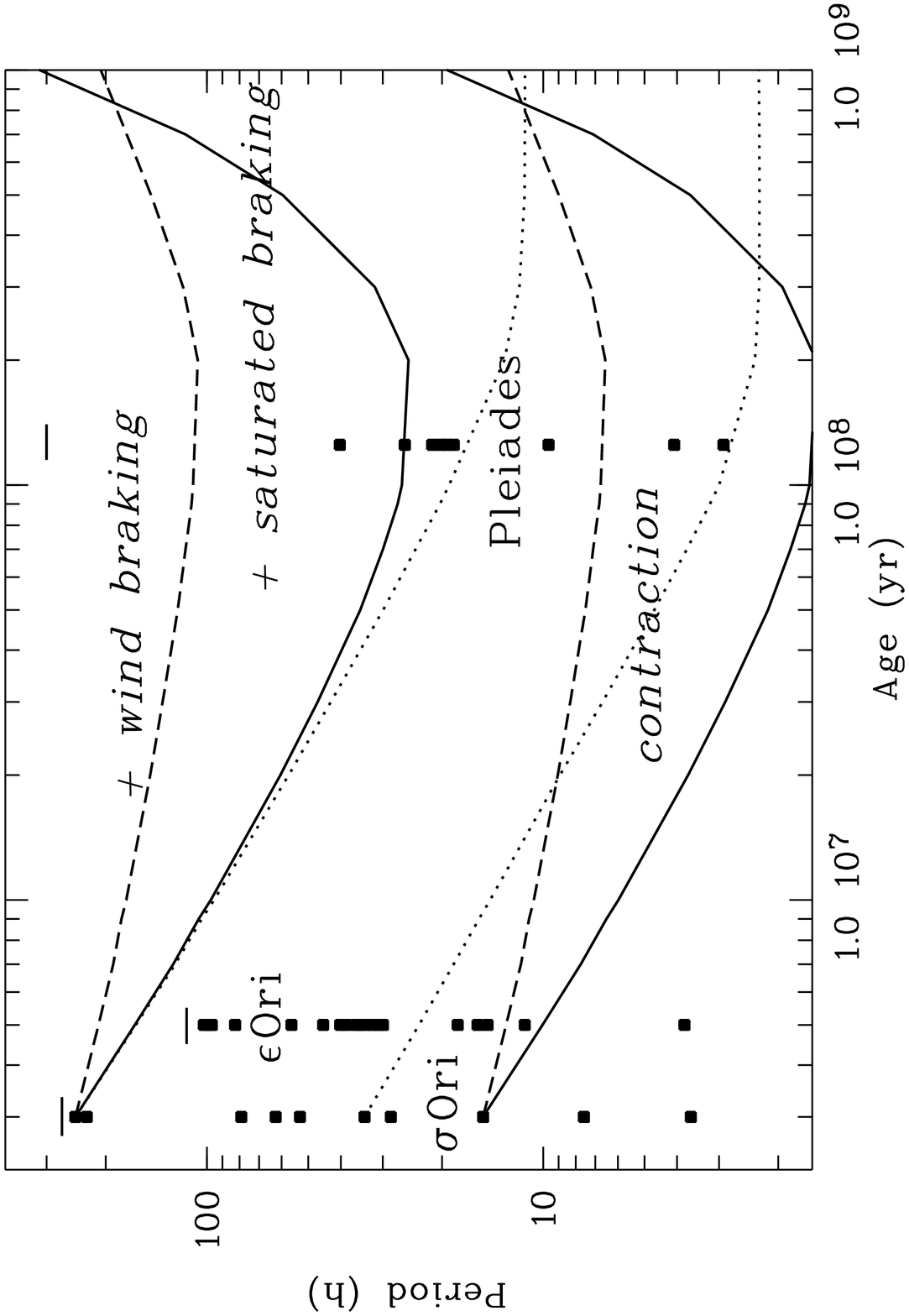}\includegraphics[angle=-90]{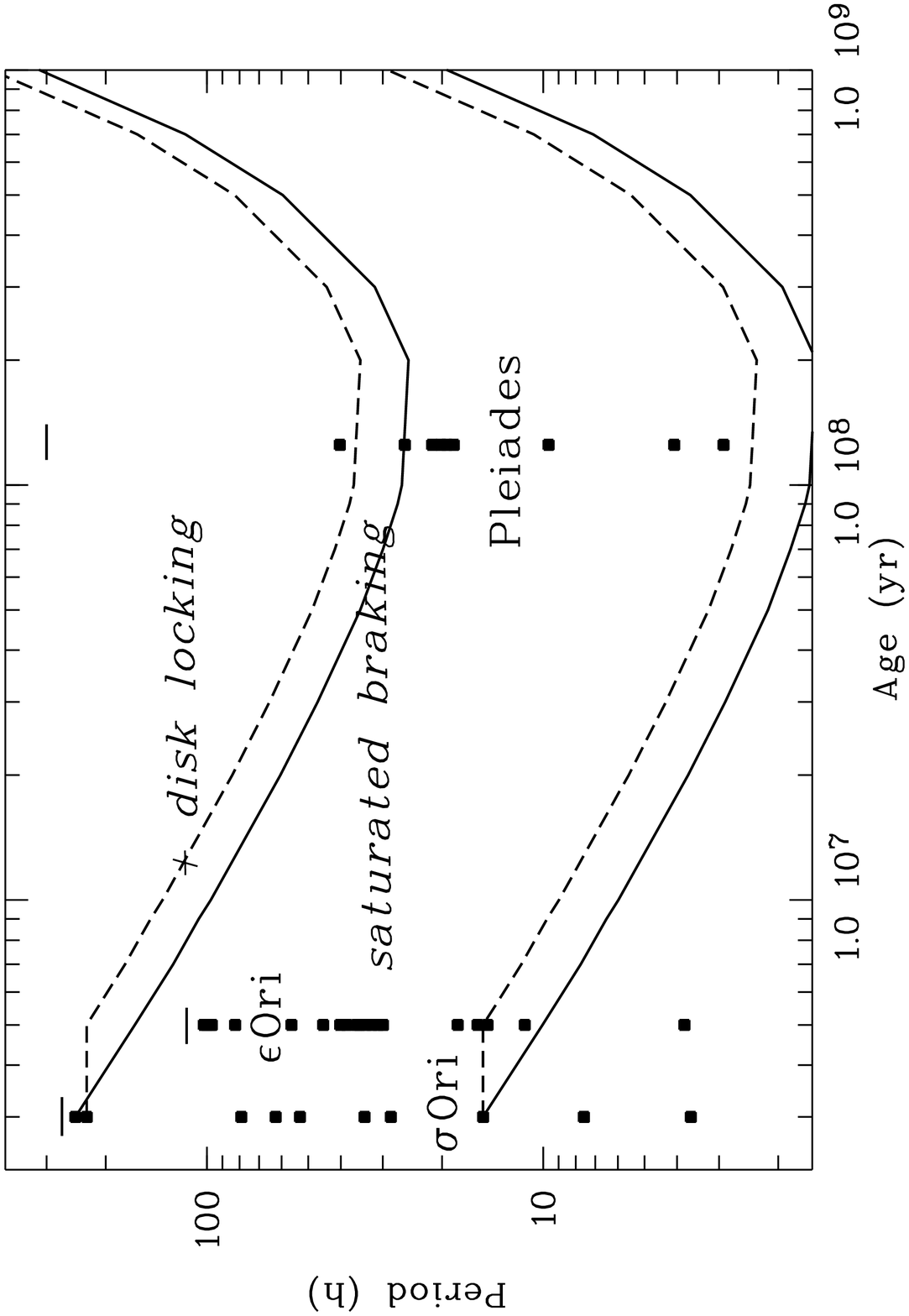}}
  \end{center}
\caption{Rotational evolution of VLM objects. The evolution of the rotation 
periods for a couple of objects for a model with hydrostatic contraction only
is shown as dotted lines. The model with additional Skumanich type wind
braking is shown as dashed lines, while saturated wind braking models are
shown as solid lines. \newline
Rotational evolution of VLM objects. The evolution of the rotation 
periods for a couple of objects for a model with hydrostatic contraction and 
saturated wind braking are shown as solid lines, as in Fig.\,4, while a model
with added disk-locking up to an age of 5\,Myr is shown as dashed lines.
\label{fig4}}
\end{figure*}

Investigating the mass dependence of the rotation periods for the VLM and
solar-mass objects in the \object{Pleiades}, we find that their period 
distributions are also different. Fig.\,3 shows that among the VLM objects we 
are lacking members with rotation periods of more than about two days, while 
the solar-mass objects show periods of up to ten days. Although our photometric
monitoring covered a time span of 18 days, we might have missed slow rotators
among the VLM objects, if their spot patterns evolved on a much shorter time
scale, or if they did not show any significant spots. In order to investigate
these possibilities, we converted the spectroscopically derived lower limits
for rotational velocities from \citet{terndrup99} and references therein
into upper limits for the rotation periods of the VLM objects using the radii
from the models by \citet{chabrier97}. These rotational velocities
should not be affected by the evolution of spot patterns on the objects. The
derived upper period limits are shown in Fig.\,3 as a solid line. With a
single exception, all our data points fall below this line, and are thus in
good agreement with the spectroscopic rotation velocities. Both complementary
data sets indicate the absence of slow rotators among the VLM objects. In
fact, our data show a trend towards faster rotation even in the VLM regime
going to lower masses. A similar trend is also seen in our \object{epsilon Ori}
sample, as well as in the Orion Nebula Cluster data by \citet{herbst01}.

\section{Rotational evolution of VLM objects}

We can now combine the periods for all three clusters, \object{sigma Ori}, 
\object{epsilon Ori}, and the \object{Pleiades} to try to reproduce their 
period distributions with simple models. These models should include essential
physics of star formation and evolution as described in Sect.\,1. Given the
currently available amount of information, we project the period distribution
for \object{sigma Ori} forward in time and comapre the model
 predictions with our observations for \object{epsilon Ori} and the
 \object{Pleiades}. 

As a first step, we take into account only the hydrostatic contraction of the
newly formed VLM objects. In this case 
the rotation periods evolve from the initial rotation period at the age of 
\object{sigma Ori} strictly following the evolution of the radii (dotted 
lines in Fig.\,3). These radii were taken from the models by 
\citet{chabrier97}. It is evident 
that this model is in conflict with the observed \object{Pleiades} rotation 
periods. Half of the \object{sigma Ori} objects would get accelerated to 
rotation periods below the fastest ones found in the \object{Pleiades} of 
about 3\,h. At the same time, even the slowest rotators in \object{sigma Ori} 
would get spun up to velocities much faster than the slower rotators in the 
\object{Pleiades}. Since the \object{sigma Ori} VLM objects surely will
undergo a significant contraction process, it is evident that significant
rotational braking must be at work until they reach the age of the 
\object{Pleiades}. 

Therefore, in a second model we add a Skumanich type braking through stellar
winds (\citealt{s72}). This wind braking acts to increase the rotation periods 
$\sim$ $t^{1/2}$, see the dashed lines in Fig.\,3. According to this model, 
some of the \object{sigma Ori} slow rotators now get braked so strongly that 
they would become clearly slower rotators than are observed in the 
\object{Pleiades} (see also Sect.\,3). This indicates that even the slowest 
\object{sigma Ori} rotators seem to rotate so fast, that they are beyond the 
saturation limit of stellar winds (\citealt{cdp95}, \citealt{tsp00}, 
\citealt{b03}). 
In this saturated regime, angular momentum loss is assumed to depend only
linearly on angular rotational velocity, thus rotation periods increase
exponentially with time. The solid lines in Fig.\,3 follow our model which
includes contraction and saturated wind braking. The period evolution of this
model clearly is the most consistent with our data.

\begin{figure}[th]
  \begin{center}
    \resizebox{\hsize}{!}{\includegraphics[angle=-90]{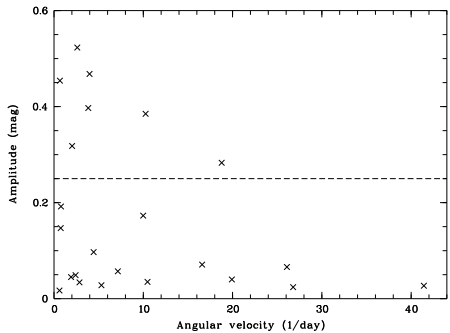}}
  \end{center}
\caption{Angular velocity versus amplitude. The dashed line delineates the 
separation between low-amplitude and high-amplitude objects.
\label{fig6}}
\end{figure}

For a few of our objects in \object{sigma Ori} we found evidence that they 
may possess an accretion disk (see below). Therefore, it is interesting 
to explore, if 
disk-locking at young age may play a role for the evolution of rotation 
periods. Assuming disk-locking for an age up to 5\,Myr, typical for the 
occurrence of accretion disks in solar-mass stars, rotation periods would 
remain constant from the age of \object{sigma Ori} (3\,Myr) to the age of 
5\,Myr. This disk-locking scenario was
combined with the saturated wind braking, with an adapted spin-down time
scale. It is shown in Fig.\,3 as dashed lines for two objects, together with
the pure saturated wind braking model discussed above (solid lines). 
The period evolution for both models is very similar. Thus from our currently
available rotation periods for these three clusters alone, there is no strong
evidence for disk-locking on VLM objects.

\begin{figure}[th]
  \begin{center}
    \resizebox{\hsize}{!}{\includegraphics[angle=-90]{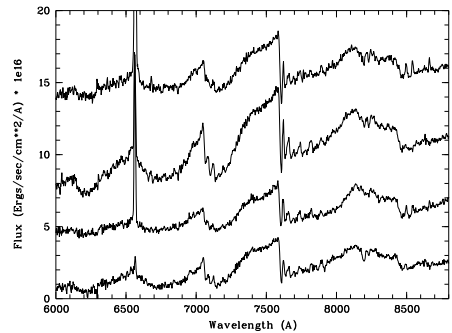}}
  \end{center}
\caption{Optical spectra for three highly variable targets (top three) and 
one non-variable target in sigma Ori. 
\label{fig8}}
\end{figure}

\section{Accretion, time variability, and disks in sigma Ori}

We note that a few of the VLM objects in the two Orion regions also show large
amplitudes of up to 0.6\,mag (see Fig.\,4). These variations are, however, of
a more irregular character and most likely result from hot spots originating
from accretion of circumstellar disk matter onto the object surface (see also
\citealt{fe96}). Optical spectra of some of these objects in sigma Ori show 
indeed emission
lines in H$\alpha$, the far-red Calcium triplet, and -- in some cases -- even
forbidden emission lines of [OI]$\lambda\lambda$6300,6363 and
[SII]]$\lambda\lambda$6716,6731, which are typical of accretion (see Fig.\,5).

It is therefore interesting to see if near-infrared excess emission, an
indicator for accretion disks, can be detected in the high-amplitude
variables. Fig.\,6 shows that non-variable and low-amplitude periodic
variables in our $\sigma$\,Ori field scatter around the isochrone for 3\,Myr
taken from \citet{baraffe98}, as expected for diskless objects in this cluster
with negligible interstellar reddening. The high-amplitude variables, on the
other hand, mostly lie in the reddening path or even red-ward of it. These
objects thus must suffer from intrinsic reddening, indicating that they indeed
possess disks. With their photometric variability, spectral accretion
signatures, and indications for near-infrared excess emission from disks
appear to be the low-mass and substellar counterparts to solar-mass T\,Tauri
stars. 

\begin{figure}[th]
  \begin{center}
    \resizebox{\hsize}{!}{\includegraphics[angle=-90]{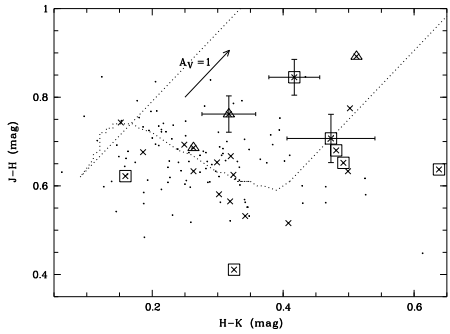}}
  \end{center}
\caption{(H-K,J-K) colour-colour diagram for sigma Ori. Crosses are 
periodically variable objects. Overplotted symbols flag objects with 
high photometric amplitude. Squares: light curve amplitude exceeds 
0.25 mag, and thus mark high-amplitude variables, triangles are 
transition objects with amplitudes between 0.1 and 0.25 mag. The 
solid line indicates the 3 Myr isochrone from the Baraffe et al. models.
Dotted lines show the interstellar extinction vector calculated from 
\citet{m90}. A reddening vector for $A_V=1$\,mag is given. 
\label{fig7}}
\end{figure}

\section{Conclusions}

We report results from our photometric monitoring of VLM objects in the
 clusters around \object{sigma Ori}, \object{epsilon Ori}, and 
the \object{Pleiades}, and first attempts to model their rotational evolution. 

VLM objects show shorter rotation periods with decreasing mass, 
which is observed already at the youngest ages, and hence must have its origin
in the earliest phases of their evolution. 

Combining the rotation periods for all our objects, we find that their
evolution does not follow hydrostatic contraction alone, but some kind of
braking mechanism, e.g. wind braking similar to the one observed in solar-mass
stars, is required as well. Such a wind braking is intimately connected to
stellar activity and magnetic dynamo action (\citealt{schatzman62}). On the 
other hand, all the 
investigated VLM objects are thought to be fully convective, and therefore
may not be able to sustain a solar-type large-scale dynamo, which is at the
heart of the Skumanich type angular momentum loss of solar-mass stars. In
fact, our modeling shows that such a Skumanich type wind braking cannot
explain our data, while saturated angular momentum loss following an
exponential braking law can. This, and the observed small photometric
amplitudes may advocate a small-scale magnetic field configuration, and may
support turbulent dynamo scenarios.

\begin{acknowledgements}
This work was partially funded by \em{Deutsche Forschungsgemeinschaft} 
(DFG), grants Ei\,409/11-1 and Ei\,409/11-2. 
\end{acknowledgements}

\bibliographystyle{aa}

\end{document}